\documentclass[aip,
 mph,%
 amsmath,
 amssymb,
 superscriptaddress,
 floatfix,
 preprint,
 eqsecnum,%
 numerical,%
]{revtex4-1}

\usepackage{graphicx}
\DeclareGraphicsExtensions{.eps,.pdf}
\usepackage[caption=false
]{subfig}
\usepackage{lineno}
\usepackage{upgreek}

\usepackage{epsfig}
\usepackage{epstopdf}
\usepackage{lipsum}
\usepackage{wrapfig}
\usepackage{wasysym}
\usepackage{threeparttable,booktabs,siunitx}
\usepackage{booktabs}
\usepackage{lmodern}
\usepackage{longtable}
\usepackage{multirow}
\usepackage{amsmath}
\usepackage[OT2,T1]{fontenc}
\usepackage[bookmarks=false]{hyperref}
\usepackage{color}
\hypersetup{pdfauthor={},pdfborder=0 0 0,colorlinks=true,citecolor=blue,linkcolor=blue,urlcolor=blue,}

\DeclareSymbolFont{cyrletters}{OT2}{wncyr}{m}{n}
\DeclareMathSymbol{\comb}{\mathalpha}{cyrletters}{"58}

\newcommand{\ben}{\begin{eqnarray}\displaystyle}
\newcommand{\een}{\end{eqnarray}}

\begin{document}
\title{Explanations of MTF discrepancy in grating-based X-ray differential phase contrast CT imaging}
\author{Yuhang Tan}
\affiliation{Research Center for Medical Artificial Intelligence, Shenzhen Institute of Advanced Technology, Chinese Academy of Sciences, Shenzhen, 518055, China}
\author{Jiecheng Yang}
\affiliation{Research Center for Medical Artificial Intelligence, Shenzhen Institute of Advanced Technology, Chinese Academy of Sciences, Shenzhen, 518055, China}
\author{Hairong Zheng}
\affiliation{Research Center for Medical Artificial Intelligence, Shenzhen Institute of Advanced Technology, Chinese Academy of Sciences, Shenzhen, 518055, China}
\affiliation{Paul C Lauterbur Research Center for Biomedical Imaging, Shenzhen Institute of Advanced Technology, Chinese Academy of Sciences, Shenzhen 518055, China}

\author{Dong Liang}
\affiliation{Research Center for Medical Artificial Intelligence, Shenzhen Institute of Advanced Technology, Chinese Academy of Sciences, Shenzhen, 518055, China}
\affiliation{Paul C Lauterbur Research Center for Biomedical Imaging, Shenzhen Institute of Advanced Technology, Chinese Academy of Sciences, Shenzhen 518055, China}

\author{Peiping Zhu}
\affiliation{Institute of High Energy Physics, Chinese Academy of Sciences, Beijing 100049, China}%

\author{Yongshuai Ge}
\thanks{Authors to whom correspondence should be addressed: Yongshuai Ge (ys.ge@siat.ac.cn).}
\affiliation{Research Center for Medical Artificial Intelligence, Shenzhen Institute of Advanced Technology, Chinese Academy of Sciences, Shenzhen, 518055, China}
\affiliation{Paul C Lauterbur Research Center for Biomedical Imaging, Shenzhen Institute of Advanced Technology, Chinese Academy of Sciences, Shenzhen 518055, China}
\begin{abstract}
As a multi-contrast X-ray computed tomography (CT) imaging system, the grating-based Talbot-Lau interferometer is able to generate the absorption contrast and differential phase contrast (DPC) images concurrently. However, experiments found that the absorption CT (ACT) images have better spatial resolution, i.e., higher modulation transfer function (MTF), than the differential phase contrast CT (DPCT) images. Until now, the root cause of such observed discrepancy has not been rigorously investigated. Through physical experiments, this study revealed that the phase grating in the Talbot-Lau interferometer induces direct superposition of paired split absorption signals and inverse superposition of paired split phase signals via diffraction. Further simulation experiments demonstrated that this splitting leads to a reduction in MTF in both ACT and DPCT images, with distinct superposition mechanisms contributing to the lower MTF in DPCT. Besides, such MTF discrepancy may also be affected in a minor extent by object composition, sample size, beam spectra and detector pixel size. Based on this study, the spatial resolution could be optimized when designing a grating-based DPC imaging system.
\end{abstract}
\maketitle
\section{Introduction}
By coupling a grating-based Talbot-Lau interferometer, the conventional X-ray computed tomography (CT) imaging system can simultaneously assess multiple contrasts\cite{Momose_2005, Weit05}, such as the absorption contrast, the differential phase contrast (DPC), and the dark-field contrast. For materials composed of light elements, the sensitivity of X-ray phase imaging may be three orders of magnitude greater than that of absorption imaging. Therefore, the phase information is particularly important in low-density object imaging, and has been considered promising during biomedical imaging.

As in many other imaging systems, the modulation transfer function (MTF) plays a crucial role in quantitatively evaluating the spatial resolution limit of a grating-based DPC imaging system. Interestingly, it was experimentally observed that DPCT shows much lower MTF than ACT\cite{Li_2013}. Despite of the plausible explanations, unfortunately, the root causes of such MTF discrepancy in grating-based X-ray DPCT imaging systems have not been rigorously investigated.

Relatedly, as the Talbot-Lau system resolution increases to nanoscale by copuling a zone plate, DPCT images have been confirmed to transition into  the superposition of paired split phase signals~\cite{Nano_DPC}. Considering the zone plate merely serves an amplifying role, the splitting phenomenon is also present in Talbot-Lau system without the zone plate. The natural question arises: does this splitting phenomenon occur in ACT images, and is it the root cause of the MTF discrepancy between ACT and DPCT.

To interpret, this study starts from the X-ray diffraction theory for the $\pi$-phase grating interferometer system incorporates with object. Theoretical derivations indicate that the detected absorption and DPC signals are both composed by a pair of diffracted signals split by a particular distance denoted as $\Delta s$, which is related to the X-ray wavelength, grating period, and the distance from the grating to detector. Specifically, the formations of the detected absorption and DPC signals are opposite: the former is generated by the direct-superposition of the paired split absorption signals, and the latter is generated by the inverse-superposition of the paired split phase signals. Because of such distinct signal superposition mechanism, we assume that different levels of edge blurring may occur on the finally reconstructed ACT and DPCT images. In other words, MTF discrepancy would appear in a grating-based X-ray imaging system. To verify, physical experiments and numerical simulation studies are performed. Additionally, object composition, sample size, beam spectra and detector pixel size are also investigated to qualify their potential impacts on such MTF discrepancy.

\section{THEORY}

\begin{figure*}[b!]
\centering
\includegraphics[width=1.0\linewidth]{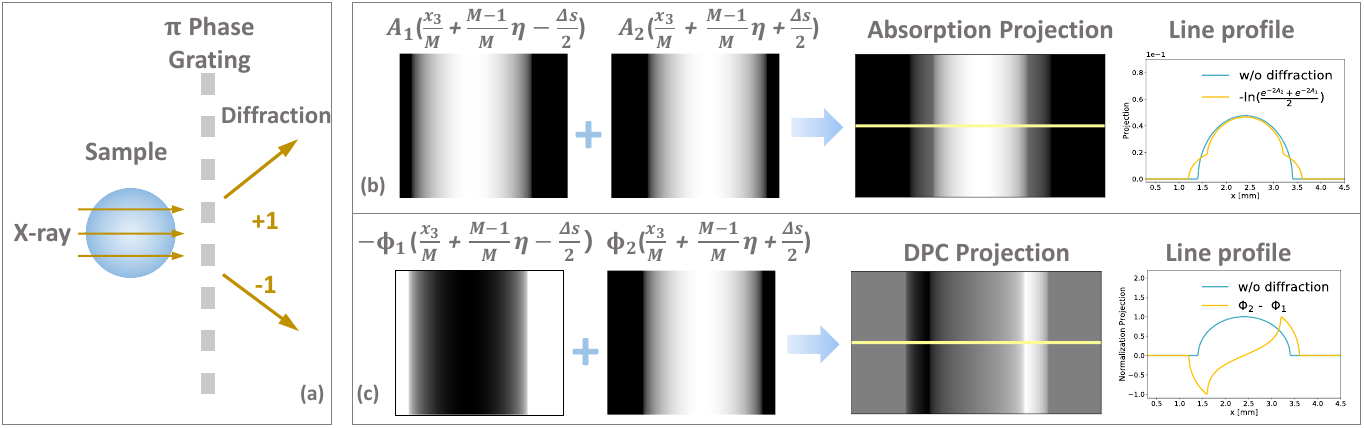}
\caption{(a) X-ray beam is diffracted after the $\uppi$ phase grating in Talbot interferometer. The -1 and +1 diffraction orders carry different sample information that are split by a distance of $\Delta s = \frac{2 \lambda d_{3}}{p_{1}}$. (b) In absorption imaging, superposition is performed on the paired positive signals. (c) In DPC imaging, superposition is performed on the paired negative and positive signals. It is clear that the absorption and DPC projections own different edge blurring effects and thus may result in MTF discrepancy.}
\label{fig:Dss_process}
\end{figure*}

In this work, the scalar wave fields of an assumed Talbot grating interferometer are calculated using the Fresnel diffraction theorem under the paraxial approximation\cite{Yong20}. In brief, the Fresnel diffraction field $\rm U_{out} (x^{\prime})$ after propagation by a distance $d$ is denoted as:
\begin{equation}
  U_{\rm{out}} (x^{\prime}) = \int U_{\rm{in}}(x) \frac{e^{ik(d + \frac{(x^{\prime}-x)^2}{2 d})}}{\sqrt{i \lambda d}} dx.
\label{eq:wave_propagation}
\end{equation}
where $U_{\rm{in}}(x)$ is the input, $\lambda$ and $k (=\frac{2\pi}{\lambda})$ represent the wavelength and wave number of X-ray, respectively. 

The scalar wave field right before the object plane, which is located $d_{1}$ distance downstream of a quasi-monochromatic point source $\delta(x_0-\eta)$, is equal to: 
\begin{equation}
U_{1} (x_{1}) = \frac{e^{ik(d_{1} + \frac{(x_{1}-\eta)^2}{2 d_{1}})}}{\sqrt{i \lambda d_{1}}},
\label{eq:wave_field_sample}
\end{equation}
where $\eta$ is the off-axis distance of the point source. Next, such wave field $U_{1} (x_{1})$ is modulated into $U_{1} (x_{1}) \cdot e^{-k\alpha(x_{1})-ik\phi(x_{1})}$ after penetrating through a thin object with a complex refractive index  $n = 1 - \delta + i \beta$. Herein, the absorption signal $\alpha(x_{1})=\int \beta(x_{1})dz$ and the phase signal $\phi(x_{1})=\int \delta(x_{1})dz$.

Immediately, the wave field on the surface of phase grating (G1) is derived as:
\begin{equation}
\begin{aligned}
&U_{2} (x_{2}) = \frac{e^{ik(d_{1} + d_{2} + \frac{(x_{2}-\eta)^2}{2 (d_{1} + d_{2})} - \Psi(\frac{d_{2}\eta + d_{1}x_{2}}{d_{1}+d_{2}}))}}{\sqrt{i \lambda (d_{1}+d_{2})}},
\end{aligned}
\label{eq:wave_field_G1}
\end{equation}
where $\Psi(x) =  \phi(x) - i \alpha(x)$. The wave field $U_{2}^{\prime}\left(x_{2}\right)$ after G1 can be expressed as follows:
\begin{equation}
U_{2}^{\prime}\left(x_{2}\right)=  U_{2}\left(x_{2}\right) T\left(x_{2}\right) = U_{2}\left(x_{2}\right) \sum_{n=-\infty}^{\infty} a_{n} e^{\frac{2 i n \pi x_{2}}{p_{1}}},
\label{eq:wave_field_after_G1}
\end{equation}
where $T(x_{2})$ denotes the periodic transmission function of G1, $p_{1}$ denotes the grating period and $a_{n}$ denotes the complex Fourier coefficient. Finally, the wave field $U_{3}\left(x_{3}\right)$ on the detection plane ($d_3$ distance away from G1) is expressed as:
\begin{equation}
\begin{aligned}
  &U_{3}\left(x_{3}\right) = \sum_{n=-\infty}^{\infty} a_{n} \frac{e^{ik(d_{1} + d_{2} + d_{3} + S(x_{3}))}}{\sqrt{i \lambda (d_{1} + d_{2} + d_{3})}},\\
  &S(x_{3})= \frac{n\lambda (d_{1} x_{3}+ d_{2} x_{3} + d_{3} \eta)}{(d_{1} + d_{2} + d_{3})p_{1}} -\frac{2n^2\pi^2d_{3}(d_{1}+d_{2})}{(d_{1} + d_{2} + d_{3})k^2{p_{1}}^2}  \\&+\frac{(x_{3} - \eta)^2}{2(d_{1} + d_{2} + d_{3})} 
-\Psi(\frac{\eta k p_{1} (d_{2} + d_{3})  + d_{1} (k p_{1} x_{3} -2 n \pi d_{3})}{(d_{1} + d_{2} + d_{3})kp_{1}}). 
\end{aligned}
\label{eq:wave_field_after_G2}
\end{equation}

Herein, we assume G1 is a $\pi$ phase grating with 0.5 duty cycle. If only considering the most dominant diffraction orders ($\pm 1$), then the detected beam intensity approximately equals:
\begin{equation}
\begin{aligned}
  I\left(x_{3}\right) &= {\left\lvert U_{3}\left(x_{3},n\right) \right\rvert}^2 \approx  \frac{{8A^2_3}}{\pi^2} (\frac{1}{2} (e^{-2A_{1}} + e^{-2A_{2}}) \\
  &+ e^{-(A_{1} + A_{2})} cos(C x_{3} + \Theta - \Phi_{1} +  \Phi_{2})),
\end{aligned}
\label{eq:intensity_on_G2}
\end{equation}
and
\begin{equation}
\begin{aligned}
  C &= \frac{4 \pi (d_{1}+ d_{2})}{(d_{1} + d_{2} + d_{3})p_{1}},\\
  \Theta &=    \frac{4 \pi d_{3} \eta}{(d_{1} + d_{2} + d_{3})p_{1}},\\
  M&=\frac{d_{1}+d_{2}+d_{3}}{d_{1}},\\
  A_{1}&=k\alpha(\frac{x_{3}}{M} - \frac{\lambda d_{3}}{M p_{1}} + \frac{M-1}{M}\eta),\\
  A_{2}&=k\alpha(\frac{x_{3}}{M} + \frac{\lambda d_{3}}{M p_{1}} + \frac{M-1}{M}\eta),\\
  \Phi_{1}&=k\phi(\frac{x_{3}}{M} - \frac{\lambda d_{3}}{M p_{1}} + \frac{M-1}{M}\eta),\\
  \Phi_{2}&=k\phi(\frac{x_{3}}{M} + \frac{\lambda d_{3}}{M p_{1}} + \frac{M-1}{M}\eta),\\
  A_{3} &=\frac{e^{ik(d_{1} + d_{2} + d_{3} +  \frac{(x_{3} - \eta)^2}{2(d_{1} + d_{2} + d_{3})})}}{\sqrt{i \lambda (d_{1} + d_{2} + d_{3})}},
\end{aligned}
\label{eq:intensity_on_G2_para}
\end{equation}
where $M$ denotes the geometric magnification, and $2\pi/C$ corresponds to the fringe period. Note that the absorption grating G2, which is usually added to generate large periodic Moir$\acute{\rm e}$ fringes, is ignored in this theoretical derivation. Eventually, the extracted absorption contrast signal ($A$) and DPC signal ($\Delta \Phi$) of the sample are:
\begin{equation}
\begin{aligned}
  &e^{-2A (x_{3})} = 0.5 \times (e^{-2A_{1}} + e^{-2A_{2}}),\\
  &\Delta \Phi (x_{3}) =  \Phi_{2}-\Phi_{1}.\\  
\end{aligned}
\label{eq:absorption_dpc_formula}
\end{equation}

Obviously, Eq.~(\ref{eq:absorption_dpc_formula}) demonstrates that the acquired absorption and DPC signals are formed via two completely different mechanisms: the former is resulted from the direct-superposition of the paired split absorption signals, and the latter is resulted from the inverse-superposition of the paired split phase signals. Intuitively, such different signal recombination mechanisms on the projections may cause different edge blurring effects and thus lead to MTF discrepancy in the reconstructed CT images, see the illustration in Fig.~\ref{fig:Dss_process} for more details.

\section{Physical experiment and results}
The physical experiments based on inverse Talbot-Lau interferometer are designed to validate the theory in Eq.~(\ref{eq:absorption_dpc_formula}). Specifically, the distance $d_0$ between source and $G_0$ is 50.0 $mm$, the distance $d_1$ between $G_0$ and $G_1$ is 177.5 $mm$, the distance $d_2$ between $G_1$ and sample is 100 $mm$, and the distance $d_3$ between sample and $G_2$ is 1701.4 $mm$. The source grating $G_0$ has a period of 2.4 $\mu m$, the phase grating $G_1$ has a period of 4.37 $\mu m$, and the  absorption grating $G_2$ has a period of 2.18 $\mu m$. The mean X-ray beam energy is 28 keV, and the detector pixel size is 4.6$\mu m$$\times$4.6$\mu m$. The sample is a gold line with a diameter of 25 $\mu m$. Under these conditions, the theoretically splitting distance $\Delta s$ is 36.5 $\mu m$, enabling the detection of the splitting phenomenon.

\begin{figure}[h]
\centering
\includegraphics[width=1.0\linewidth]{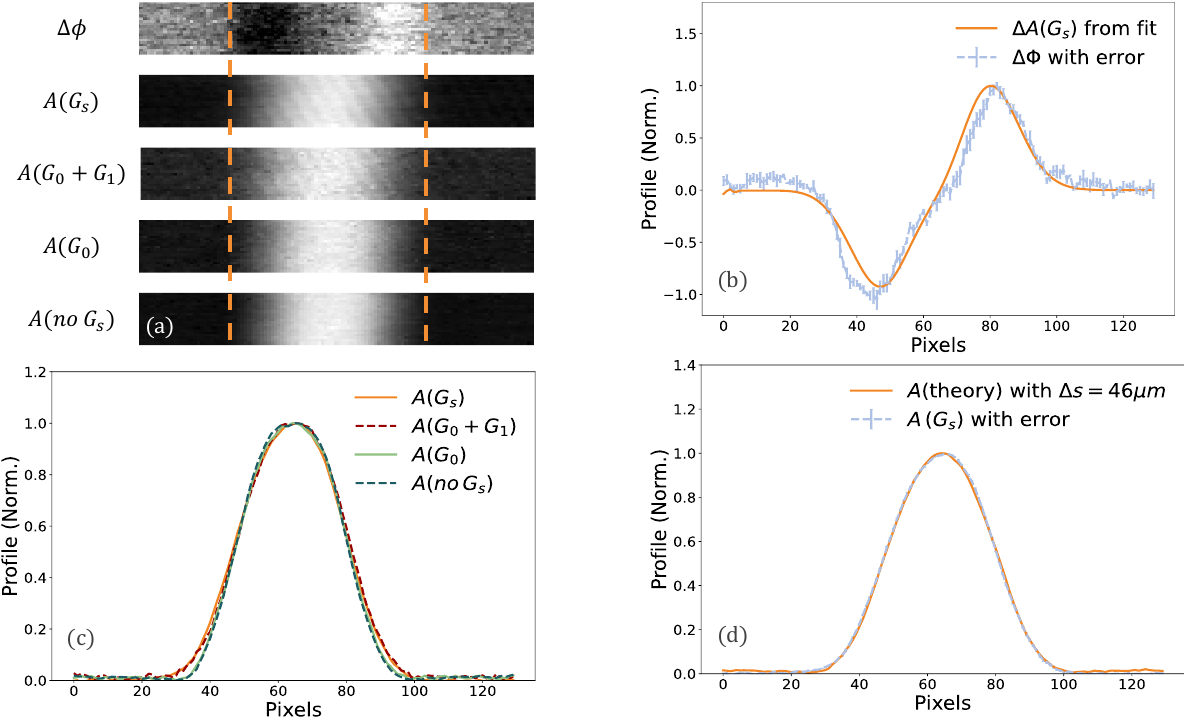}
\caption{(a) The phase and absorption images of the gold line sample with different grating combinations. Comparison of vertically averaged profiles: (b) $\Delta A (G_s)$ and $\Delta \phi$; (c) Different grating combinations; (d)$ A (G_s)$ and $A(\text{theory})$ with $\Delta s = 46\mu m$. }
\label{fig:experiment_re}
\end{figure}

To verify, the experiments employ different grating combinations: \(G_s=G_0+G_1+G_2\), \(G_0+G_1\), \(G_0\), and \(no \, G_s\), and the phase and absorption images are shown in Fig.~\ref{fig:experiment_re}(a). In addition, the vertically average profiles are compared. Specifically, the sample width of $\Delta A(G_s)$ obtained by fitting to reduce noise is the same as $\Delta \phi(G_s)$, as shown in Fig.~\ref{fig:experiment_re}(b). The sample width in the absorption image with $G_1$ (e.g. $A(G_s)$,$A(G_0+G_1)$) are larger than that without $G_1$ (e.g. $A(G_0)$,$A(no G_s)$), as shown in Fig.~\ref{fig:experiment_re}(c). When $A(G_0)$ or $A(no G_s)$ is substituted into Eq.~(\ref{eq:absorption_dpc_formula}) with $\Delta s = 46\mu m$, its sample width becomes consistent with $A(G_0+G_1)$ or $A(G_s)$, aligning with the theoretical expectation with an error of about two pixels, as shown in Fig.~\ref{fig:experiment_re}(d). These physical experiment results verified that the splitting phenomenon also exists in the ACT image. However, due to factors such as excessive system length and low X-ray energy, each projection takes nearly a day to acquire, making it impossible to obtain experimental MTF results. Therefore, simulations are emploied to explore the root case of the MTF discrepancy between ACT and DPCT.

\section{Simulation experiments and results}
The numerical simulation platform was developed in Python (Version: 3.9). The propagation of wave fields was implemented via the fast Fourier transform (FFT). More details of this numerical simulation platform can be found in \cite{Yang_2022}.

\begin{figure*}[h]
  \centering
  \includegraphics[width=1.0\linewidth]{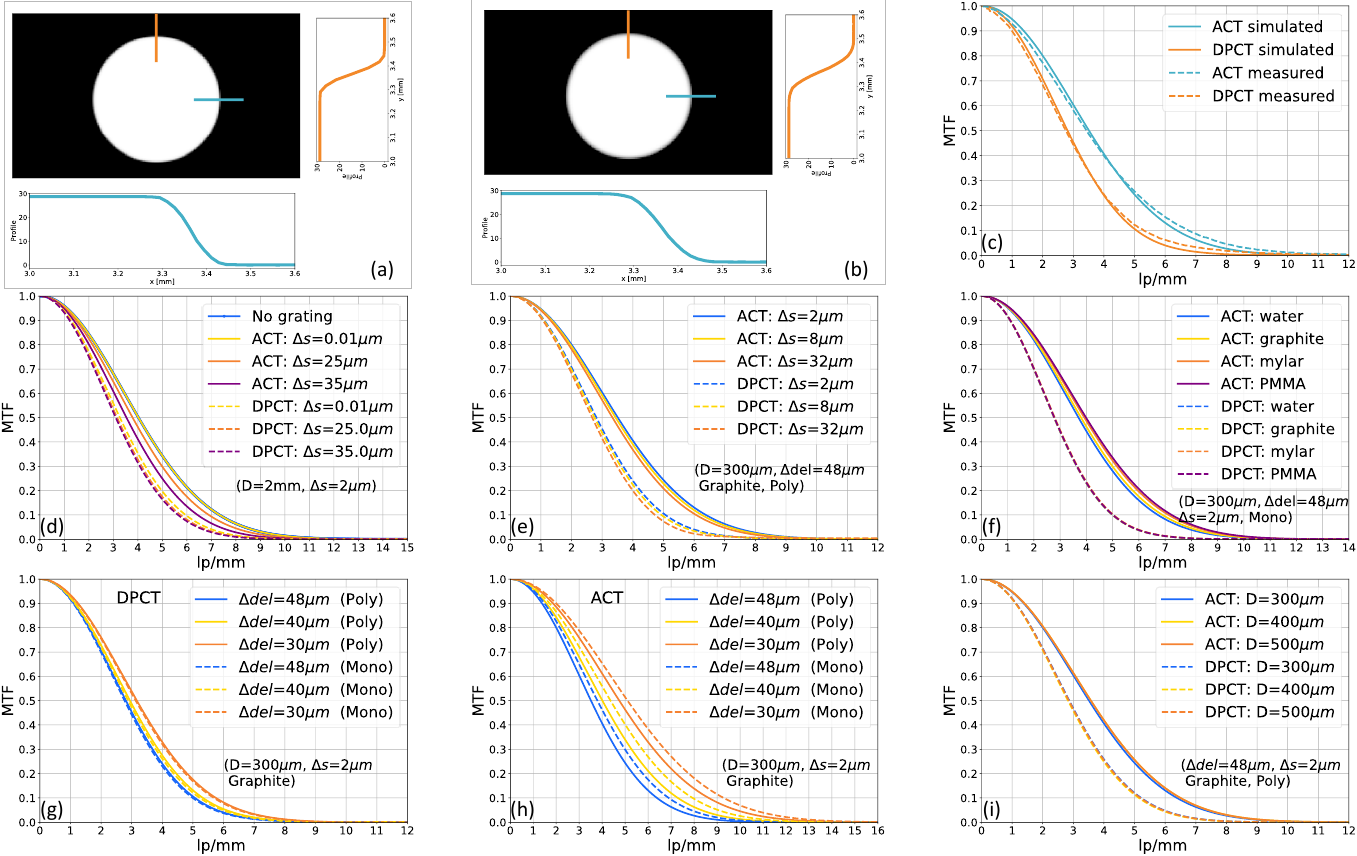}
  \caption{(a) and (b) depict the reconstructed ACT and DPCT images of the graphite, respectively. Edge profiles along both horizontal and vertical directions are plotted. (c)Comparison results of measured MTFs and simulated MTFs of ACT and DPCT. (d) Theoretically estimated MTF curves for different $\Delta s$ according to Eq.~(\ref{eq:absorption_dpc_formula}). (e) Numerically simulated MTF responses for different $\Delta s$. (f) Dependence of MTF curves on object compositions. (g) MTFs of DPCT for different detector pixel size $\Delta del$ and beam spectra. (h) MTFs of ACT for different $\Delta del$ and beam spectra. (i) MTF responses for different object size $D$. Note that other imaging parameters are listed inside each plot.} 
  \label{fig:impact_MTF}
  \end{figure*}

The first simulation employs the same system configuration as used in \cite{Li_2013}. By doing so, the viability of the theoretical assumption can be validated. The ACT and DPCT images with 10$\mu \rm m$$\times$10$\mu \rm m$ pixel size were reconstructed from the filtered-back-projection (FBP) algorithms\cite{DPCCT_method}, see Fig.~\ref{fig:impact_MTF}(a) and (b). In particular, the Ramp filter was used for ACT reconstruction, and the Hilbert filter was used for DPCT reconstruction\cite{ACT_method}. Results are plotted in Fig.~\ref{fig:impact_MTF}(c). As seen, the simulated and measured MTF curves of ACT and DPCT exhibit high agreements. This demonstrates the feasibility of our provided explanation: the inverse-superposition in phase imaging degrades the image spatial resolution more significantly than the direct-superposition in absorption imaging, provided that the standard ACT and DPCT image reconstruction algorithms are implemented.


Additional simulations were performed to investigate the dependency on $\Delta s$, object composition, beam spectra, sample size (D) and detector pixel size ($\Delta$del), correspondingly. Results are shown in Fig.~\ref{fig:impact_MTF}. Plots in Fig.~\ref{fig:impact_MTF}(d) are numerically calculated from Eq.~(\ref{eq:absorption_dpc_formula}) for different $\Delta s$. Undoubtedly, results demonstrate that the different formation mechanisms of absorption signal and DPC signal are the primary cause of MTF discrepancy. The simulated results are plotted in Fig.~\ref{fig:impact_MTF}(e). It is found that the MTFs of both ACT and DPCT get lower as $\Delta s$ increases. This is understandable since the edge blurring effect gets more severe. It is worth mentioning that the adjustment of $\Delta s$ is achieved by varying the period of G1, and $\Delta s < \Delta del$ is assumed.

  \begin{table}[htbp]
    \centering
    \footnotesize
    \caption{\bf Results of MTF for different materials.}
    \begin{tabular}{ccccc}
    \hline
      & water & graphite & mylar & PMMA  \\
      \hline
      $\delta$($\times10^{-7}$) &3.69&3.90 & 4.84&4.26 \\
      $\beta$($\times10^{-11}$) &10.48& 9.39 &9.14&7.45 \\
      $\delta / \beta$ &3521&4153&5295&5718  \\
    $\rm MTF^{ACT}_{10\%}$ [lp/mm] &6.7&6.9&7.1&7.2   \\
    $\rm MTF^{DPCT}_{10\%}$ [lp/mm]&5.0&5.0&5.0&5.0  \\
      \hline
    \end{tabular}
      \label{tab:six_materials}
    \end{table}
    
As listed in Table.~\ref{tab:six_materials}, four different materials are simulated. Results are depicted in Fig.~\ref{fig:impact_MTF}(f). Interestingly, the MTFs of ACT get higher as the ratio $\delta / \beta$ increases, whereas, the MTFs of DPCT are approximately the same for different object compositions.

The energy dependent results of the MTFs are plotted in Fig.~\ref{fig:impact_MTF}(g) and Fig.~\ref{fig:impact_MTF}(h) along with varied detector pixel sizes. As seen, the MTF increases as the pixel size reduces. For DPCT, the impact of beam spectra is negligible. As a contrary, the ACT shows slightly better MTF performance with monochromatic 25 keV X-ray beams. The polychromatic X-ray beam is simulated under 40 kVp with 1.0 mm Al filtration. It is noticed that the sample size has little impact on the MTF of DPCT, see the results  in Fig.~\ref{fig:impact_MTF}(i). The MTF of ACT becomes slightly better as the object size increases.
  
\section{Discussions and CONCLUSION}
This paper provides a feasible explanation for the MTF discrepancy of ACT and DPCT acquired from the same grating-based interferometer system, and the different superposition of the projection signals might be the main cause. Physical experiment results show that the detected absorption projection follows the direct-superposition of the paired split absorption signals, and the detected DPC projection follows the inverse-superposition of the paired split phase signals. Numerical simulation results show that the inverse-superposition is more easy to degrade the edge sharpness. Therefore, the MTF of DPCT is lower than that of ACT. Numerical simulations demonstrate the high agreement with the previous experimental measurements. Besides, the diffraction induced splitting distance $\Delta s$, object composition, beam spectra, sample size and detector pixel size are further investigated. Results find that the MTFs of both ACT and DPCT would get lower as $\Delta s$ increases. However, the ACT always outperforms the DPCT. Additionally, the object composition, sample size, beam spectra and detector pixel size may also affect the MTF.

Based on this study, we think additional attention is required when discussing the spatial resolution in different X-ray interferometer systems and imaging tasks, particularly for ACT and DPCT. Note that the current simulations are implemented for small signal splitting $\Delta s$, i.e., within one detector pixel. For systems having larger $\Delta s$ (or small detector pixel dimension), however, innovative signal extraction techniques\cite{Nano_DPC,Takano_DPC_Nano} need to be considered. Essentially, more rigorous ACT and DPCT image reconstruction algorithms are desired to generate high precision ACT and DPCT images for any X-ray interferometer system. To do so, we think the derived projection signal models in Eq.~(\ref{eq:absorption_dpc_formula}) would be helpful in initiating such future studies. Finally, the MTF discrepancy effect would be evaluated for the $0.5\pi$-phase grating interferometer system\cite{Momose_2005} and other none grating-based DPC imaging approaches such as the diffraction enhanced imaging\cite{Diffraction_enhancement}.

In conclusion, a feasible explanation for the MTF discrepancy phenomenon of ACT and DPCT in grating-based X-ray interferometer system is investigated in this study. Upon it, the performance of the spatial resolution of a given multi-contrast grating-based imaging system can be optimized.

\section{Funding} 
This project is supported by the National Natural Science Foundation of China (12027812), Guangdong Basic and Applied Basic Research Foundation (2021A1515111031), and the Youth
Innovation Promotion Association of the Chinese Academy of Sciences (2021362).

\section{Disclosures} 
The authors declare no conflicts of interest.

\section{Data availability} 
Data underlying the results presented in this paper are not publicly available at this time but may be obtained from the authors upon reasonable request.

\bibliography{sample}

\end{document}